\documentclass[10pt,conference]{IEEEtran}
\usepackage{blindtext}

\usepackage{amsmath}
\usepackage[utf8]{inputenc}
\usepackage{amsmath} 
\usepackage{amssymb}
\usepackage{bigstrut}

\usepackage{graphicx}
\usepackage[framed,bw]{mcode}
\usepackage{multirow}
\usepackage{url}

\newcommand{\Fig}[1]{Fig.~\ref{#1}}
\newcommand{\eq}[1]{(\ref{#1})}
\newcommand{\Tab}[1]{Tab.~\ref{#1}}
\newcommand{\Sec}[1]{Sec.~\ref{#1}}

\newcommand{\E}[1]{\mathop{\bf E}\nolimits\left[#1\right]} % Expectation was rm before

\newcommand{\eps}{\varepsilon}
\newcommand{\ns}{\mbox{\textit{ns}-3}}
\newcommand{\DopplerSp}{\nu_d}

\graphicspath{{./images/}}

\begin{document}

%\title{Performance Analysis of a LTE Downlink Scheduler Providing Fair Throughput Guarantees}
%\title{Fair Throughput Guarantee Scheduler for LTE: When and Where}
\title{Scheduling Policies for the  LTE Downlink Channel: A Performance Comparison}

\author{
	\IEEEauthorblockN{
		Mattia Carpin\IEEEauthorrefmark{1}, Andrea Zanella\IEEEauthorrefmark{1}, Jawad Rasool\IEEEauthorrefmark{2}, Kashif Mahmood\IEEEauthorrefmark{3}, Ole Gr\o ndalen\IEEEauthorrefmark{3}, Olav N. \O sterb\o\IEEEauthorrefmark{3}
	}       % <-this % stops a space
	\IEEEauthorblockA{
		\IEEEauthorrefmark{1} Dept. of Information Engineering, University of Padova,~Italy\\
		\IEEEauthorrefmark{2} Telenor,~Norway  \\
		\IEEEauthorrefmark{3} Telenor Research,~Norway  
	}
}

% The paper headers
\markboth{Here the header of the Page (if any)}%
{Shell \MakeLowercase{\textit{et al.}}: Bare Demo of IEEEtran.cls for Journals}

\maketitle

\begin{abstract}
A key feature of the packet scheduler in LTE system is that it can allocate resources both in the time and frequency domain. Furthermore, the scheduler is acquainted with channel state information periodically reported by user equipments either in an aggregate form for the whole downlink channel, or distinguished for each available subchannel. This mechanism allows for wide discretion in resource allocation, thus promoting the flourishing of several scheduling algorithms, with different purposes. It is therefore of great interest to compare the performance of such algorithms in different scenarios. A very common simulation tool that can be used for this purpose is $\ns$, which already supports a set of well known scheduling algorithms for LTE downlink,
% such as maximum throughput, and blind equal throughput. However, the simulator
though it still lacks schedulers that provide throughput guarantees. In this work we contribute to fill this gap by implementing a scheduling algorithm that provides long-term throughput guarantees to the different users, while opportunistically exploiting the instantaneous channel fluctuations to increase the cell capacity. 
%The algorithm, which was previously proposed to work in the time domain only, is here extended to the frequency domain, under the assumption that the downlink channel is affected by frequency selective Rayleigh fading. 
We then perform a thorough performance analysis of the different scheduling algorithms by means of extensive $\ns$ simulations, both for saturated UDP and TCP traffic sources. The analysis makes it possible to appreciate the difference among the scheduling algorithms, and to assess the performance gain, both in terms of cell capacity and packet service time, obtained by allowing the schedulers to work on the frequency domain.

\end{abstract}

\begin{IEEEkeywords}
Resource management, opportunistic scheduling, LTE networks, $\ns$, fair throughput guarantee.
\end{IEEEkeywords}

\IEEEpeerreviewmaketitle

\section{Introduction}
\label{section:introduction}
Long Term Evolution (LTE) with its IP based flat network architecture promises improved spectral efficiency and high data rate to its users.
As of first quarter of 2014, LTE users have already reached the 200 million mark with mobile data traffic in Q1 2014 exceeding the total mobile data traffic in 2011~\cite{EricsonMobiltyReport}.
It is reported in \cite{EricsonMobiltyReport} that there will be 2.6 billion LTE subscriptions by the end of 2019, while a 10x growth in mobile data traffic is predicted between 2013 and 2019.

In order to address this challenge an efficient radio resource management module is needed of which the packet scheduler is an important component.
The downlink packet scheduler at the medium access control (MAC) layer is indeed in charge of dynamically allocating the downlink radio resources to the user equipments (UEs), thus determining the order of service and the transmit rate of each user. 

%the orthogonal frequency division multiple access~(OFDMA) based downlink in LTE allows the packet scheduler to make use of both the time dimension (allocation of time frames) and the frequency dimension (allocation of sub-carriers)

One of the key features of LTE is that it allows resource allocation both in the time and the frequency domain. The scheduler can hence decide to allocate all the resources in a given time interval to a single user, or partition them also in the frequency domain, assigning fraction of the available bandwidth to different users in the same time slot. As a result, the scheduler can choose between a simple allocation policy where a single user gets all the resources in a given time slot (TD), and a more sophisticated approach where resources are allocated with a finer granularity, also exploiting the frequency dimension (FD). The scheduler, furthermore, can make use of the channel quality indication (CQI) that is periodically reported by each UE, either in an aggregate form for the whole downlink channel, or distinguished for each available subchannel. While TD schedulers only need the aggregate CQI, the potential of FD schedulers is fully available when the CQI is provided on each subchannel. The increased flexibility of FD schedulers in resource allocation, however, is paid back in terms of higher complexity and signaling overhead. 
%  traditionally the FD approach makes use of the \texttt{subband cqi}, which is channel quality value for a specific subband, while the TD approach makes use of the \texttt{wideband cqi} which is an estimate of the channel quality over the entire wideband.
It is therefore of utmost importance to investigate the scenarios where the FD implementation of the schedulers brings significant improvements over the simpler TD version. 

%The downlink packet scheduler at the medium access control (MAC) layer dynamically allocates the downlink radio resources to the user equipments (UEs).
%It is a key component that affects the behavior of LTE systems.
The LTE standard does not impose any restriction on the type of scheduler, thus leaving space for innovation. The main challenge in the design of an LTE scheduler is to find a proper balance between partially contrasting objectives. 
%Further this \emph{open} nature of the packet scheduler allows a lot of innovation.
On the one hand, in fact, resource allocation shall increase the spectral efficiency and, in turn, the cell capacity, which is a key performance index from the operator perspective. On the other hand, resource allocation shall also consider user-related constraints, such as fairness and the Quality of Service~(QoS) requirements. This makes the design of the packet scheduler a challenging optimization problem, which has been addressed in different ways \cite{BET}.

%The MAC scheduler allocates the downlink radio resources to the UEs according to a priority metric which varies for the different scheduling algorithms.
%A compilation of the priority metrics for the different scheduling algorithms can be found in \cite{survey}.

% This calls for innovative solutions to meet the growing traffic demand.
%In its first release, LTE supported a downlink peak rate of 300 Mbit/s which shows a considerable increase in the spectral efficiency when compared to the previous cellular systems.

For example, the \emph{Maximum Throughput Scheduler} (MTS), also known as the \emph{opportunistic scheduler}, prioritizes the cell capacity by exploiting channel variations among UEs, while the \emph{Blind Equal Throughput Scheduler} (BETS) aims to provide throughput fairness among the users irrespective of their channel quality, thus at the cost of cell capacity. 
\emph{Proportional Fair Scheduler} (PFS) somehow balances the channel awareness and throughput fairness aspects of MTS and BETS, respectively, by assigning resources proportionally to the average channel quality of each UE. Finally there is also a rich class of schedulers, such as the token bank fair queue scheduler, the aim of which is to provide QoS guarantees \cite{art:TBFQ}.

The aforementioned schedulers along with some others with somewhat similar priority metrics are already implemented in $\ns$, which is a widely used network simulator~\cite{ns3}. To the best of our knowledge the state of the art on LTE schedulers falls short when it comes to schedulers which provide throughput guarantees to users. In this work, we contribute to fill this gap by implementing in $\ns$ the \emph{Fair Throughput Guarantees Scheduler} (FTGS), which was
% and carrying out a performance comparison of the same with the existing LTE schedulers implemented in $\ns$. FTGS, 
originally proposed in \cite{opportunistic} in TD mode, and that is designed to guarantee equal \emph{long-term} throughput to all UEs, while opportunistically exploiting the temporal variability of the downlink channels to increase the cell capacity. Here we extend FTGS to work in FD mode, and compare the two versions of the algorithms with other schedulers for LTE systems already supported by $\ns$, both for flat and frequency selective fading channels. 

%matched to the LTE structure, thereby enhancing the existing $\ns$ repository of QoS aware schedulers. 

%We carry out a performance comparison of the TD version of FTGS with the respective version of MT, BETS and PFS under flat fading channel.
The analysis reveals that FTGS provides a good trade-off between cell capacity and fairness in different scenarios, both for saturated UDP and TCP traffic sources. Furthermore, we observe that the FD version of FTGS can bring substantial improvements over TD in frequency selective channels, when the channel dispersion is large, thereby justifying the increased computational complexity brought by FD implementation.
In addition, we analyze the inter-scheduling time at MAC layer of FTGS, BETS, and PFS, in both fast and slow fading scenarios, in order to assess the potential impact of such scheduling algorithms on delay-sensitive applications. The analysis shows that the inter-scheduling time of the TD version of FTGS can be indeed critical in presence of slow fading channels. However, we observe that the FD version of FTGS can dramatically reduce the inter-scheduling time and the service time of high-layer packets in case of frequency selective channels, thus alleviating the aforementioned problem. 

%As it is not straight forward to relate the inter-scheduling delay at the MAC layer with the inter-scheduling delay at TCP layer we propose a simple method to carry out the TCP delay analysis.
%We find out that for FTGS the average packet delay relative to a fixed byte count remains constant. This is because of the inherent working principle of FTGS which is to schedule good users rarely and vice versa.
%The effect of the different SINR distributions, which maps to the UEs locations in the cell, on the cell spectral efficiency is then quantified showing the proposed algorithm performs poorly in terms of cell efficiency if equal throughput guarantees are promised to all users. 

In summary, following are the main contributions of this paper:
\begin{itemize}
\item We propose an extension of the FTGS scheduler for the FD mode that makes it possible a finer distribution of the transmission resources to the UEs, thus potentially improving the spectral efficiency. As ancillary contribution, we enrich the existing $\ns$ repository with schedulers that support long term fairness and throughput guarantees to the UEs.\footnote{The software can be downloaded from \url{http://goo.gl/3BYxOC}.} 
\item We investigate and compare the performance of many different schedulers for LTE downlink channel, both in the TD and FD versions.
\item The performance analysis is carried our for different types of fading channels, and both for saturated UDP and TCP traffic sources.  Furthermore, service time statistic are also analyzed and compared for different schedulers, when varying the fading characteristics. 
\item We analyze the opportunistic gain of the FTGS, with BETS being the benchmark, for the case when the users signal to interference noise ratios are disparate.
\item Finally, the strength of the FTGS scheduler has been tested against the impact of an imprecise channel estimation. 
\end{itemize}

The remainder of this paper is organized as follows. Sec.~\ref{RW} briefly summarizes the related work. We describe the system model, simulation scenario and performance metric in Sec.~\ref{SM}. Sec.~\ref{SP} focuses on various scheduling policies considered in this paper. Sec.~\ref{section:simulation} describes the simulation setup, while Sec.~\ref{section:results} shows the numerical results. Finally, conclusions and future works are presented in Sec.~\ref{section:conclusions}.

\section{Related work}\label{RW}
Performance analysis of downlink scheduler can be carried out either at the MAC layer or at the transport layer where in the latter the effect on transmission control protocol (TCP) is of high importance.
While the MAC throughput analysis has its own benefits, a huge fraction of today's data is carried via HTTP~\cite{TCPUsage} which uses TCP because it is reliable, well understood, and can be conveniently managed by firewalls and security systems.
However, it is not alway straightforward to infer TCP throughput from MAC performance for a given scheduling algorithm.
It is therefore essential to investigate the performance of the scheduling algorithms when they handle TCP traffic as well. 

Acknowledging this need, the performance analysis presented in \cite{implementation}, which compared the MAC-layer throughput of some schedulers available in $\ns$, has been extended in \cite{evaluation} to TCP traffic sources, comparing both the aggregate and per-user TCP throughput achieved by the different schedulers. 
%In \cite{evaluation}, the authors present a performance analysis, in terms of cell and individual user throughput for TCP traffic, of some of the scheduling algorithms currently available in \emph{ns}-3. 
%In fact the MAC layer analysis in \cite{implementation} is extended to TCP analysis in \cite{evaluation}.
%Recently there has been a renewed interest in the TCP analysis of the scheduling algorithms and hence designing algorithms which take into account the TCP traffic as opposed to the UDP traffic.
%For example 
In \cite{aware} a TCP-aware scheduling algorithm, named Queue MW, has been implemented in $\ns$. The performance of Queue MW is then compared against other scheduling policies in terms of throughput and delay for different queue sizes at the evolved node B (eNodeB).
Similarly the adverse effect, due to the variability in inter-scheduling time brought by PFS on TCP and its congestion control mechanism is highlighted in \cite{enhanced}.
It needs to be mentioned that PFS is commonly used as a reference scheduler for many performance analysis study. For example the fairness and bit rate characteristics of MTS and PFS are compared in \cite{PFS_MU_LTE_Kwan09}.
A PFS based scheduling algorithm is proposed in \cite{FreeAndMultiUserDiv_PFS_LTE_Mobility_PIMRC_12} which takes into account the frequency diversity and the multi-user diversity gain simultaneously.
The algorithm is tested by simulating in a heterogeneous environment consisting of both high-mobility and low mobility users and with saturated sources. The MAC analysis reveals substantial improvements in overall cell throughput as compared to raw PFS.
Finally the effect of the chosen scheduling algorithm on cell spectral efficiency and packet delay experienced by the user for VoIP and video traffic is analyzed in \cite{LTE_Downlink_PerformAnaly:Biernacki_2014}.

In a nutshell, there has been a growing interest in the design and performance comparison of the scheduling algorithms for LTE taking into account both the UDP and TCP traffic. However, the comparison with opportunistic but users' fair schedulers, capable of providing throughput guarantees to users while exploiting the specific resource allocation structure of LTE, has not been carried out.
Secondly, although several well-known scheduling algorithms are already implemented in $\ns$ \cite{implementation,evaluation,aware} but to the best of our knowledge, QoS aware schedulers which provide fair throughput guarantees to the users and take into account the LTE resource allocation framework have not yet been implemented in $\ns$.
This paper aims at filling these gaps.

\section{System model}\label{SM}
In this section we first recall the reference architecture of LTE system, which is necessary to understand the working principle of the packet scheduler, and then describe the framework considered in our work. 

\subsection{LTE basics}\label{section:frame_structure}
\begin{figure}[t]
	\centering		 \includegraphics[width=0.70\columnwidth]{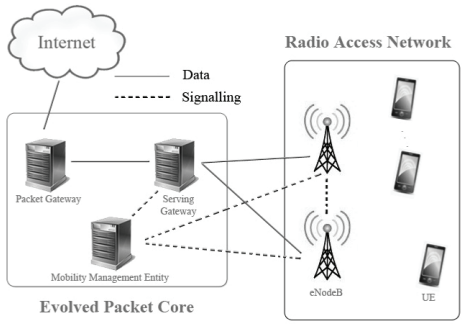}
	\caption{LTE high level architecture.}
	\label{fig:LTEArchitecture}
\end{figure}

The high-level architecture of LTE system is sketched in \Fig{fig:LTEArchitecture}. %, which is necessary to understand the working principle of the packet scheduler.
There are two keys components: the radio access network (RAN) and the evolved packet core (EPC).
The RAN provides the wireless connectivity between the eNodeB and the UE while the EPC is responsible, among other things, for connecting the RAN to the Internet for end-to-end communication.
%In this work we use the LTE model in $\ns$~\cite{ns3} to create a LTE network which involves all the core elements such as EPC.
The resource allocation in LTE is done in a centralized manner and as such both the uplink and the downlink scheduler reside inside the eNodeB. 

%The resources are the physical \emph{Resource Blocks} (RBs), each of which corresponding to a time-frequency slot, whose bit rate and capacity are determined by the modulation and coding scheme used in that RB. 
%
%The downlink packet scheduler dynamically allocates the downlink RBs to the user equipments (UEs). In case of semi persistent scheduling, the resource allocation remains valid for multiple Transmission Time Intervals (TTIs). In this work, however, we focus on the non persistent scheduling, where resource allocation is repeat at each TTI. 
%\AZ{The following sentence is not very clear to me.... instead, we shall mention that the scheduler knows the channel state information of the UE at the beginning of each TTI. We shall also define the TTI... }
%Here it is pertinent to mention that apart from the resource management the control procedures are also handled by the eNodeB. 

LTE uses an air interface based on Orthogonal Frequency Division Multiple Access (OFDMA) for downlink while for the uplink the Single Carrier Frequency Division Multiple Access (SC-FDMA) is used~\cite{LTEsesia}. In this paper we focus on the downlink side, where OFDMA is employed. 

\begin{figure}[t]
	\centering		 \includegraphics[width=\columnwidth]{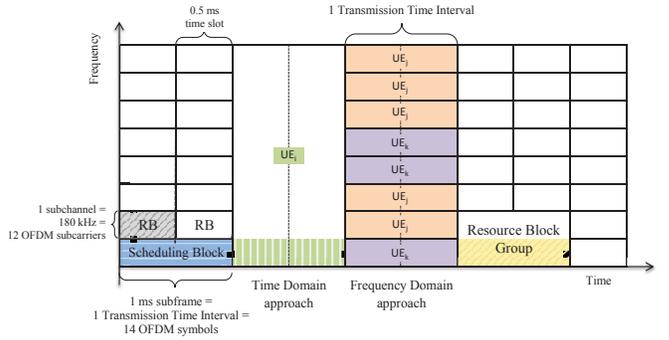}
	\caption{LTE frame structure.}
	\label{fig:LTEFrame}
\end{figure}

The frame structure of the downlink air interface is shown in \Fig{fig:LTEFrame}. Each frame consists of 10 subframes of 1ms duration, with each subframe further divided into two slots of $0.5$ms each.
A slot consists of 7 OFDM symbols in the time domain (normal cyclic prefix), and is divided in the frequency domain into a number of Resource Blocks (RBs) depending on the available channel bandwidth. The bit rate and capacity of each RB are determined by the modulation and coding scheme (MCS) used in that RB. It needs to be mentioned that LTE standard imposes a restriction in that all the RB's assigned to the same user in a given transmission time interval (TTI) must use the same MCS. The smallest radio resource unit that can be assigned by the scheduler to an UE is the scheduling block which is two adjacent RBs in TD and one sub-channel in FD. From here onwards the Resource Block (RB) refers to the scheduling block and the time slot refers to the length of the subframe (1 TTI).

A Resource Block Group (RBG) consists of multiple adjacent RBs in a single time slot. 
In each 1 ms subframe, the MAC scheduler is responsible for allocating the RBs to one or more UEs according to the specific scheduling metric, and the TD or FD approach. 
In this work we consider {non-persistent} scheduling, where the resource allocation is repeated at each subframe as opposed to the {semi persistent} scheduling for which the resource allocation remains valid for multiple subframes.

%In each 1 ms subframe the MAC scheduler is in charge for allocating the RBs to each UE according to the specific scheduling metric. In the FD approach, the MAC scheduler distinguishes the resources along both the time and frequency domain. Therefore it allocates resources per RB-base achieving the finest possible granularity. In the TD approach, instead, all the RBs in the same subframe are assigned to the same user. 

Resource Allocation Type specifies the way in which the scheduler allocates RBs for each transmission. The actual version of $\ns$ supports only Allocation Type 0, where first the scheduler divides RBs in RBG, with a number of RBs in each group that depends on the system bandwidth. Then, each RBG is assigned to a UE according to the scheduling metric. For example, if we consider an overall cell bandwidth of 25 RBs, each RBG contains 2 RBs \cite{LTEsesia}, thus the scheduler must assign for each time slot the 12 available RBGs to a user according to the scheduling metric, leaving the last RB unscheduled.

When performing the scheduling decision, the MAC scheduler can make use of the CQI reported by each UE and used by the radio resource management to estimate the channel's quality \cite{LTEsesia}. There are two possible estimations that each UE can perform: a wideband estimation, where a single CQI value is reported for the entire bandwidth, and the subband estimation, where the CQI is evaluated and reported for each RB. We assume that the TD scheduling approach makes use of the wideband CQI, while the FD implementation can access the information on the subband CQI.

\subsection{Spectral efficiency}
Denoting by $\gamma$ the SINR of an UE, its spectral efficiency can be expressed as
\begin{equation}
\eta=\log_2 \Big(1+\frac{\gamma}{\Gamma} \Big),
\label{eta}
\end{equation}
where $\Gamma$, sometimes referred to as $\rm SNR_{\rm gap}$ \cite{Gamma}, accounts for the difference between the theoretical Shannon bound and the efficiency obtained by practical modulations \cite{inproc:SINRgap}. For the downlink channel of LTE, $\Gamma$ can be obtained from the target Bit Error Rate (BER) as ~\cite{Gamma}
\begin{equation}
\Gamma=-\frac{\ln \big(5 \cdot BER\big)}{1.5}.
\end{equation}
The spectral efficiency is then mapped to the CQI according to Tab.~\ref{CQIeta}, where $\eta_{th}$ defines the upper boundary of the interval within which CQI value does not change. We assume that each UE reports this CQI value to the eNodeB before the scheduling decision takes place \cite{3GPP_Table}. It should be noted that a higher CQI value corresponds to a better channel.
\begin{figure}[t]
	\centering		 \includegraphics[width=0.95\columnwidth]{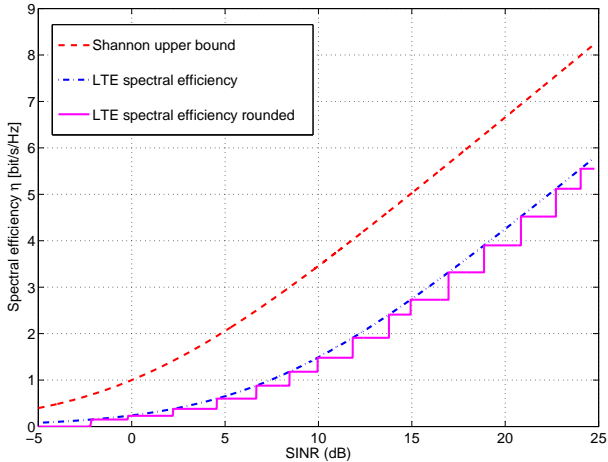}
	\caption{Spectral efficiencies as a function of the SINR.}
	\label{fig:eta}
\end{figure}
\begin{table}[htbp]
  \centering
  \caption{Spectral efficiency and CQI mapping.}
    \begin{tabular}{c|cccccccc}
    CQI   & 0     & 1     & 2     & 3     & 4     & 5     & 6     & 7 \\
    \hline
    $\eta_{th}$ & 0.15  & 0.23  & 0.38  & 0.6   & 0.88  & 1.18  & 1.48 & 1.91\\
    \hline \hline
    CQI   & 8     & 9     & 10    & 11    & 12    & 13    & 14    & 15 \\
    \hline
    $\eta_{th}$ & 2.41  & 2.73  & 3.32  & 3.9   & 4.52  & 5.12  & 5.55 & \textgreater 5.55\\
    \end{tabular}
  \label{CQIeta}%
\end{table}

\Fig{fig:eta} shows the difference between the spectral efficiencies computed using Shannon upper bound ($\Gamma = 1$) and that given by \eq{eta} for a target $\mathrm{BER} = 5 \cdot 10^{-5}$ ($\Gamma \simeq 5.53$). We also plot the discrete version of the spectral efficiency, obtained from Tab.~\ref{CQIeta}. The $\ns$ implementation of the LTE network makes use of this version according to the standard specifications \cite{3GPP_Table}. For the analysis that will follow, however, we assume the difference between the continuous and discrete spectral efficiencies to be negligible. 

%\section{System model}
%\label{section:system_model}
\subsection{Simulation scenario}
We consider a system in which a single eNodeB serves a population of $N$ backlogged users where  $\bar{G_i}$ is the average rate experienced by user $i$ when it is scheduled.
We assume that eNodeB receives the CQIs from all UEs before a scheduling decision is performed. The eNodeB is also assumed to know the Signal to Interference-plus-Noise Ratio (SINR) distribution of each user exactly.
However we also show later the effect of imperfect channel estimation on FTGS.
This estimation of SINR distribution can, for example, be performed from the CQI feedback coming from different users in the system.

Within a single cell, LTE networks provide orthogonality among users, both in the downlink and uplink. This means that different users, served by the same eNodeB, are assigned different resources in order to avoid interference among them. In this work, since we assume a single eNodeB and a single cell, we are implicitly neglecting the presence of interference. However, since the analysis and the scheduling algorithm can still be applied in the presence of interference, we use the term SINR even for the single cell scenario. 

Finally, we consider a path loss channel model affected by either flat or frequency selective fading, and by additive white Gaussian noise.

\subsection{Performance Metrics}
The purpose of this study is to compare the different scheduling policies for the downlink channel of an LTE system, when varying the channel conditions and the number of users. As mentioned, each scheduling algorithm offers a different balance between the overall spectral efficiency of the cell, and the service offered to each user. Therefore, the comparison will be performed considering a set of performance indexes, both at the MAC and transport layer. More specifically, we will analyze the \emph{cell throughput}, defined as the aggregate throughput obtained by all the UEs in the cell, together with the Jain's fairness index $J$, defined as~\cite{jain}
\begin{equation}
J=\frac{\big(\sum_{i=1}^{N}x_i\big)^2}{N \sum_{i=1}^{N}x_i^2}.
\end{equation}
where $x_i$ is the throughput achieved by the $i$th UE.
It should be noted that $J=1/N$ corresponds to the minimum fairness, while $J=1$ indicates perfect fairness among the users in the system. In addition, we consider the statistical distribution of the inter-scheduling time, that is the time a UE is not scheduled for transmission. From this metric, we then estimate the mean and standard deviation of higher-layer packet service time, which is relevant for delay-sensitive applications.

\section{Scheduling policies}\label{SP}

In this section we describe the scheduling policies considered in our analysis. We start by describing the TD version of the schedulers, where all RBGs are allocated to a single UE at each time slot. Successively, we extend the analysis to the FD case, where RBGs in the same time slot can be allocated to different UEs. 

\subsection{TD versions of the scheduler}
The scheduling decision in the time domain decides which UE will get all the RBGs in the upcoming slot $k$. The decision is based on a priority metric that varies for the different scheduling algorithms. In the following, we describe the metric of the schedulers considered in this study. 

%In the FD version, multiple UEs can receive data within a single time slot since each RBG $l$ in the time slot $k$ is allocated independently, whereas in TD version 
\subsubsection{Maximum Throughput Scheduler (MTS)}
%\paragraph*{Maximum Throughput Scheduler (MTS)}
As discussed earlier, opportunistic schedulers exploit instantaneous channel variations to maximize the cell throughput. One such algorithm is MTS that schedules the users with more favorable channel conditions. The associated scheduling metric is then
\begin{equation}
i_{\!_{MTS}}(k)=\operatorname*{arg\,max}_{1 \leq i \leq N} r_i(k),
\label{eq:TDMT}
\end{equation}
where $r_i(k)$ is the instantaneous rate of user $i$ in time slot $k$, and it depends on the wideband CQI.  It is worth remarking that MTS can achieve the maximum cell throughput but, if the average SINR distributions of different users are extremely unbalanced, it could result in the starvation of UEs experiencing bad channel conditions. 

\subsubsection{Blind Equal Throughput scheduler (BETS)} On the other hand, BETS guarantees equal throughput among all users in the system. The scheduling metric for BETS is defined as follows~\cite{BET}
\begin{equation}
i_{\!_{BETS}}(k)=\operatorname*{arg\,max}_{1 \leq i \leq N} \frac{1}{\zeta_i(k)},
\label{eq:BET}
 \end{equation}
where ${\zeta_i(k)}$ is the past average throughput of the $i$-th UE at time slot $k$, which is given by (see \cite{LTEsesia})
\begin{equation}
{\zeta_i(k)}=\beta \cdot \zeta_i(k-1)+ (1-\beta) \cdot r_i(k),
\label{betsS}
\end{equation}
where $\beta \in [0,1]$.
It should be noted that BETS is a channel-unaware scheduler, and thus it is not very efficient in terms of cell throughput. Due to its simplicity, the long-term throughput achieved by BETS can actually be computed in a closed form, as explained in Appendix~\ref{BETS}. 

\subsubsection{Proportional Fair scheduler (PFS)} PFS can improve the cell throughput by incorporating channel conditions in the BETS. This can be observed in the following scheduling metric of the PFS, where current rate is considered in addition to the past throughput:
\begin{equation}
i_{\!_{PFS}}(k)=\operatorname*{arg\,max}_{1 \leq i \leq N}  \frac{r_i(k)}{\zeta_i(k)} \,.
\label{eq:PF}
\end{equation}

\subsubsection{Fair Throughput Guarantees scheduler (FTGS)} 
In the literature, many priority based opportunistic schedulers have been proposed, where different priorities are given to UEs based on certain fairness criteria. FTGS is one such scheduler, which was originally proposed in \cite{opportunistic} in the TD version. The scheduling metric of the FTGS is given as 
\begin{equation}
i_{\!_{FTGS}}(k)=\operatorname*{arg\,max}_{1 \leq i \leq N}  \frac{r_i(k)}{\alpha_i}, 
\label{metric}
\end{equation}
where $\alpha_i$ is a constant assigned to user $i$, which is chosen such that the scheduler maximizes the throughput guarantees of all UEs within a time window $T_W$. These constants depend on the SINR distributions of the users, such that a user with poor average channel conditions, i.e., lower average SINR, is given higher priority compared to users with better average channel conditions. It should be noted that, when all the users have equal average SINR distributions, $\alpha_i=\alpha$ $ \forall i$, then the FTGS reduces to MTS.

The set of coefficients $\{\alpha_i\}$ in \eqref{metric} is obtained by solving an optimization problem. We report in Appendix~\ref{Optimization} the derivation of the coefficients, along the same lines as \cite{opportunistic}, but taking into account the practical aspects of LTE system related to resource allocation.

%Choosing a good starting point to solve the above system of non linear equations is crucial to guarantee a fast convergence to the optimum solution. We argue that users with low $\bar{\gamma}$ should be given higher access probability. To this end we set
%$p_0(i)=1/\bar{\gamma}_{i,dB}$ and, then, normalize the initial access probabilities so that $\sum_{i}p_0(i)=1$. Finally, we choose $R_{i,0}=0.1\cdot \bar{\gamma}_{i,dB}$ and $\alpha_{i,0}=0.5 \cdot \bar{\gamma}_{i,dB}$.{\color{red} I have commented before that we should remove this information since it is an internal step we took to solve this.} {\color{blue} I may agree with Jawad. Maybe we could say that if we have already solved the optimization problem for a particular SINR distribution, and we have a variation on the average SINR of one user, we can use the previous solution as initial point to provide a faster convergence.} \AZ{I totally agree with Jawad and suggest to remove this last paragraph.}

\subsection{FD generalization of the schedulers}
So far, we have assumed that all the RBGs in a slot have to be assigned to a single UE. We here consider the possibility of scheduling multiple UEs in the same time slot by assigning one single RBG at a time. We then denote by $r_i(k,l)$ the rate that user $i$ can get from RBG $l$ at slot $k$. 

The FD versions of the MTS and PFS can then be obtained by changing the priority indexes as follows:
\begin{equation}
\hat{i}_{_{MTS}}(k,l)=\operatorname*{arg\,max}_{1 \leq i \leq N}  r_i(k,l)
\label{eq:FDMT}
\end{equation}
\begin{equation}
\hat{i}_{_{PFS}}(k,l)=\operatorname*{arg\,max}_{1 \leq i \leq N}  \frac{r_i(k,l)}{\zeta_i(k)} 
\label{eq:PF}
\end{equation}

The FD version of BETS is based on the same principle of the TD version, that is providing equal throughput to all UEs. The only difference is that RBGs are allocated one at a time, and the throughput of each UE shall be updated accordingly. More specifically, the UE with the lowest past average throughput is selected, and assigned the first RBG. The expected throughput of this UE is then calculated using \eq{betsS}, but replacing $r_i(k)$ with $r_i(k)/M$, where $M$ is the number of RBGs that can be separately allocated in that time slot. %is now estimated as the fraction of the rate user $i$ can achieve by using \emph{all RBGs} in that time slot, 
The new throughput is compared with the average throughput $\zeta_i(k)$ of the other UEs. The scheduler keeps allocating RBGs to the same UE until its expected throughput is no longer the smallest. The strategy is then applied to the new UE with the lowest $\zeta_i(k)$, until all RBGs are allocated. 

For what concerns FTGS, we observe that the rationale to derive the coefficients $\{\alpha_i\}$ can be extended to the frequency domain. This is of particular interest when a frequency selective channel is considered. In this context, making use of the subband CQI, the eNodeB can guarantee more fair scheduling while improving the cell throughput. If we take the channel to be frequency selective, with each subband assumed to be narrow enough to be considered frequency-flat, the SINR is still exponentially distributed, and the $\alpha_i$ values obtained in Tab.~\ref{tab:solution} are valid and can be reused. Otherwise, the optimization problem can be solved again according to the new SINR distribution and a new set of $\alpha_i$'s can be obtained. The FTGS scheduling metric for the FD approach becomes
\begin{equation}
\hat{i}_{\!_{FTGS}}(k,l)=\operatorname*{arg\,max}_{1 \leq i \leq N}  \frac{r_i(k,l)}{\alpha_i} ,
\label{metricFD}
\end{equation}

The FD implementation achieves higher granularity at the cost of higher implementation complexity. It is, therefore, important to investigate the trade-off between system improvement due to FD approach and increased computational complexity and signaling cost. This complexity not only comes from the scheduler, in the eNodeB, that needs to provide flexible allocation in the frequency domain, but also from the UE, that needs to measure the CQIs for each subband, instead of reporting a single value for the entire bandwidth.

We argue that the improvements brought by the FD implementation depend on the frequency selectivity of the channel, that is, the channel dispersion. A parameter that is normally used to define the channel dispersion is the root-mean square (rms) delay spread, $\tau_{\rm rms}$, which corresponds to the second-order central moment of the channel impulse response \cite{benvenuto}, that is
\begin{equation}
\tau_{\rm rms}=\sqrt{\big(\bar{\tau}^2\big)-\big( \bar{\tau} \big)^2},
\end{equation}
with
\begin{equation}
\bar{\tau}^n=\dfrac{\sum_{i=1}^N \E{|g_i|^2}\tau_i^n}{\sum_{i=1}^N \E{|g_i|^2}},
\end{equation}
where $\tau_i$ is the delay of the $i$-th subband, and $\E{|g_i|^2}$ is its average statistical power with $\E{.}$ denoting the expectation.

We then expect that the higher $\tau_{\rm rms}$, the larger the gain of the FD schedulers over their TD counterparts, whereas for low values of $\tau_{\rm rms}$ (almost flat channels), the two approaches are expected to exhibit similar performance. To confirm our intuition, both the TD and FD implementations will be tested on channels with different dispersion indexes.

\section{Simulation Setup}
\label{section:simulation}
%\subsection{System parameters}
\begin{table}[t]
  \centering
    \caption{System parameters setting.}
    \begin{tabular}{|l|l|}
    \hline
    \textbf{Parameter} & \textbf{Value} \\
    \hline
    Number of RBs & 25 \\
    Bandwidth & 5 MHz \\
    RBG size & 2 \\
    Downlink EARFCN & 500 \\
    AMC Model   & Piro \\
    Pathloss model & Friis \\
    Fading model & Trace based \\
    Error mode control & Deactivated \\
    Radio Link Control Mode & Unacknoledged \\
    Tx power of eNode & 30 dBm \\
    Tx power of UEs & 23 dBm \\
    Noise figure at eNode and UEs & 5dB \\
    Transmission Time Interval   & 1ms \\
    TCP packet size & 1024 B \\
    \hline
	\end{tabular}%
  \label{tab:parameters}%
\end{table}%
To assess the performance of the schedulers in different environments, the Network Simulator version 3.20 ($\ns$) has been used, which is, at the time of writing, the latest release.  

A default EUTRA-Absolute Radio Frequency Channel Number (EARFCN) of 500 is used, that corresponds to a carrier frequency of $f_c=2.16$ GHz. We use the unacknowledged mode for the radio link control (RLC) layer and the adaptive modulation and coding (AMC) model proposed in \cite{LTEAMC} for $\ns$. Various system parameters are summarized in Tab.~\ref{tab:parameters} while the ones not reported have the default value which comes in standard $\ns$ setting. 

The wireless link is modelled as a path loss plus fading channel. The $\ns$ LTE module includes a trace-based fading model that makes use of pre-calculated traces to limit the computational complexity of the simulations \cite{ns3}. All users share the same fading trace but with random starting point in order to have almost independent fading processes.
We analyze the behaviour of different scheduling policies both for the flat fading and frequency selective channels, where the traces can be obtained by using MATLAB script that comes with the $\ns$ release. We assume that the channel temporal correlation follows Jakes' model, and we denote with $\DopplerSp$ the Doppler spread. The users' speeds can be modified to simulate fast and slow fading environments where, as we will see later in the paper, the delay experienced by a packet is subject to great variations. Frequency selective Rayleigh channels have been generated as proposed by the 3GPP~\cite{3GPP_pdp}. More specifically, we considered the power delay profiles reported in Tab.~\ref{tab:pdp}, which refer to pedestrian, vehicular, and urban environments, with $\tau_{\rm rms}$ being 44~ns, 356~ns, and 990~ns, respectively. %The corresponding coherence bandwidth, computed as $1/ \tau_{\rm rms}$, is then  22.7 MHz, 2.8 MHz and 1MHz. 
%We expect the FD version of FTGS to improve the performance in the second and third scenarios, while performing similarly to the TD version in the pedestrian environment.

\begin{table}[t]
  \centering
  \caption{Parameters for the frequency selective channels.}
    \begin{tabular}{c|c||c|c||c|c}
    \multicolumn{2}{c||}{Pedestrian, $\tau_{\rm rms}=44$~ns} & \multicolumn{2}{c||}{Vehicular, $\tau_{\rm rms}=256$~ns} & \multicolumn{2}{c}{Urban, $\tau_{\rm rms}=990$~ns} \\
    \hline
    $\tau_i$ (ns) & $E[|g_i^2|]$ (dB) & $\tau_i$ (ns) & $E[|g_i^2|]$ (dB) & $\tau_i$ (ns) & $E[|g_i^2|]$ (dB) \\
    \hline
    0     & 0.0   & 0     & 0.0   & 0     & -1.0 \\
    30    & -1.0  & 30    & -1.5  & 50    & -1.0 \\
    70    & -2.0  & 150   & -1.4  & 120   & -1.0 \\
    90    & -3.0  & 310   & -3.6  & 200   & 0.0 \\
    120   & -8.0  & 370   & -0.6  & 230   & 0.0 \\
    190   & -17.2 & 710   & -9.1  & 500   & 0.0 \\
    410   & -20.8 & 1090  & -7.0  & 1600  & -3.0 \\
          &       & 1730  & -12.0 & 2300  & -5.0 \\
          &       & 2510  & -16.9 & 5000  & -7.0 \\
          \hline
    \end{tabular}%
  \label{tab:pdp}%
\end{table}%

Unless otherwise specified, simulations are carried out by considering $N=10$ static UEs, with average SNR values $\{\bar \gamma_i\}$ as reported in the first column of Tab.~\ref{tab:solution} (the values in the other columns will be described later). 
The linear mean of such values is here referred to as \emph{mean cell SINR}, which is given by
%define the cell average SINR as %{\color{red} Should we instead say "average of users' average SINR" instead of overall average?} {\color{blue} I propose "cell average SINR"}
\begin{equation}
\mu_{dB}=10 \log_{10} \Bigg(\frac{1}{N} \sum_{i=1}^{N}\bar{\gamma}_{i} \Bigg).
\label{eq:averageSINR}
\end{equation}
It may be worth remarking that $\mu_{dB}$ depends on the position of the UEs with respect to the center of the cell, and it will be used in the following to compare scenarios with different UEs' location. With reference to the values in Tab.~ \ref{tab:solution}, the mean cell SINR turns out to be $\mu_{dB}=15$~dB. 

%In the first column of Tab.~ \ref{tab:solution}, the average SINR of each user is reported, with $\mu_{dB}=15$ dB and the $\bar{\gamma}$ of the worst user taken as 10 dB. 
Simulations have been carried out by considering both saturated UDP traffic sources (that is saturated traffic at the MAC layer) and saturated TCP sources that generate traffic toward each UE. Each simulation lasts for 60 seconds of simulated time, enough to average out the fading fluctuations and achieving an excellent statistical confidence.\footnote{Since the $95\%$ confidence interval is generally very narrow, it has been omitted from the figures to reduce clutter.} %, and has been repeated for FTGS has been compared with MTS, PFS and BETS. 

\begin{table}[t]
  \centering
  \caption{Average SINR and FTGS parameters of users.}
    \begin{tabular}{c||c|cccc}
    i     & $\bar{\gamma}_{i,dB}$ & $\alpha_i$& $p(i)$  & $\bar{R}_i/W$ \\
    \hline
    1     & 10.0000   & 2.9899& 0.1490 & 2.5114 \\
    2     & 11.7041  & 3.7867 & 0.1235  & 3.0292 \\
    3     & 12.9248  & 4.3845& 0.1099  & 3.4031 \\
    4     & 13.8766 & 4.8634 & 0.1012 & 3.6950 \\
    5     & 14.6568 & 5.2634& 0.0951  & 3.9342 \\
    6     & 15.3180  & 5.6070& 0.0904  & 4.1365 \\
    7     & 15.8917   & 5.9083& 0.0868 & 4.3117 \\
    8     & 16.3984  & 6.1768& 0.0838  & 4.4662 \\
    9     & 16.8521   & 6.4190 & 0.0812& 4.6043 \\
    10    & 17.2628  & 6.6397& 0.0791  & 4.7291 \\
    \hline
    \end{tabular}%
  \label{tab:solution}%
\end{table}%

\section{Numerical results}
\label{section:results}
In this section we present the simulation results for the TD and FD version of the four schedulers under different fading conditions. We initially consider a flat fading channel, in which TD and FD approaches perform exactly the same, because of the lack of diversity in the frequency dimension. Therefore, we present results for TD schedulers only. Successively, we consider different frequency selective channels, and simulate both the TD and FD
versions of the schedulers. %The goal is to see whether the increased computational cost brought by the FD version, is compensated by a significant improvement in the system performance.

\subsection{Flat fading channel}
To begin with, we analyze the parameters used by the FTGS scheduler in this scenario, which are reported in the second and third column of Tab.~\ref{tab:solution}, while the rightmost column gives the expected rate of each UE when it gets scheduled, as for \eq{shannon}, normalized to the channel bandwidth $W$. We see that, as expected, the access probability $p(i)$ is lower for users with better channel conditions, that is larger $\bar \gamma_i$, which are then scheduled more rarely in order to leave more resources to users with bad channel. 
The average spectral efficiency of the $i$-th UE can be obtained as
\begin{equation}
\bar{\eta}=  \frac{p(i) \cdot \bar{R}_i}{W}=0.374\,,
\end{equation}
which is equal for all users, since we assume identical throughput guarantees. At the same time, FTGS shall be able to increase the cell efficiency by opportunistically exploiting the channel variations in the short term. 
%\AZ{The following paragraph (in green) is not really connected with the analysis we are carrying on. I suggest to remove it.}
%{\color{green} Anyhow, the optimization problem that defines the FTGS strategy can be rewritten in order to differentiate the throughput guarantees to the users \cite{opportunistic}. This can be useful if we wish to provide different guarantees according to the channel conditions, or a different service to users belonging to different classes. }

\begin{figure}[t]
	\centering		 \includegraphics[width=\columnwidth]{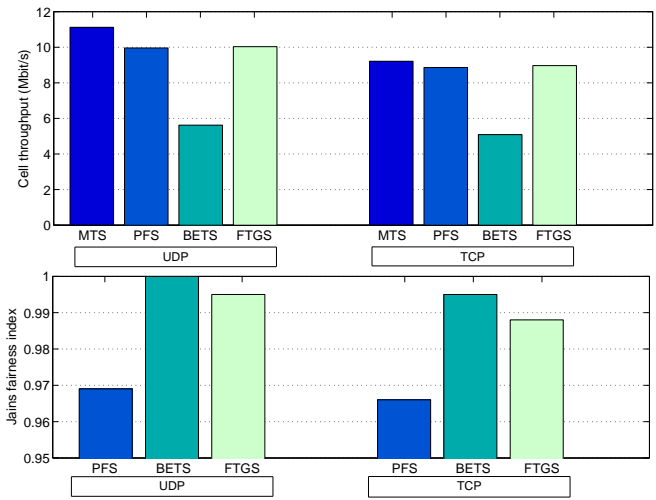}
	\caption{Average cell rate (Mbit/s) and Jain's fairness index achieved by the four considered schedulers, using a flat fast fading channel ($\DopplerSp=120$ Hz).}
	\label{fig:rate}
\end{figure}

To investigate these properties, we report in \Fig{fig:rate} the throughput and the fairness achieved by the TD version of the different schedulers, both for UDP and TCP saturated traffic, considering a flat fast-fading channel with Doppler spread $\DopplerSp=120$ Hz.

Note that, to better appreciate the comparison between the fairness performance of PFS, BETS, and FTGS, we removed from the lower plot of \Fig{fig:rate} the results for MTS, whose Jain's fairness index is significantly lower than that of the other algorithms, being approximately equal to $0.62$ and $0.82$ for the MAC and TCP cases, respectively. 

From \Fig{fig:rate} we can observe that, as expected, the opportunistic nature of MTS yields the best results in terms of aggregate cell throughput, both for UDP and TCP traffic. 
Conversely, the channel-agnostic approach of BETS yields the highest fairness in both scenarios, but the overall cell throughput is considerably reduced. FTGS and PFS, instead, perform fairly well both in
terms of throughput and fairness, with an apparently small advantage of FTGS over PFS. 

\begin{figure}[t]
	\centering		 \includegraphics[width=\columnwidth]{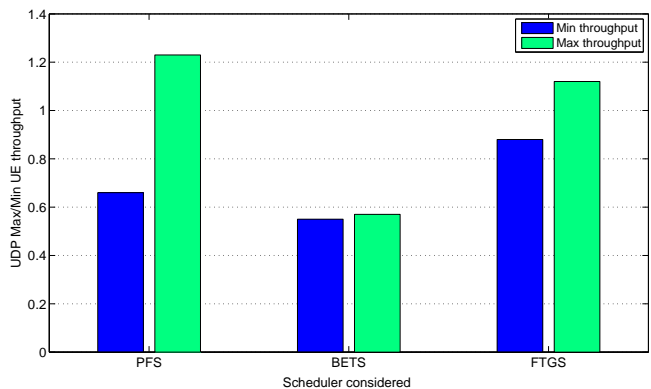}
	\caption{Minimum and Maximum average throughput per UE achieved by the 4 considered schedulers, using a flat fast fading channel ($\DopplerSp=120$ Hz) and an UDP saturated traffic.}
	\label{fig:minmax}
\end{figure}

From the fairness plot in \Fig{fig:rate}, we can note that the fairness obtained with TCP traffic is slightly reduced as compared to UDP traffic. This is because an UE is not considered for scheduling until the eNodeB receives an acknowledgement for the previous data from this UE. The population of users is therefore not constant over time for the TCP case. Compared to the MAC-saturated scenario, FTGS is thus not able to provide perfect fairness to all users.

To gain more insight on the performance of the schedulers (PFS, BETS and FTGS), we report in \Fig{fig:minmax} the worst and the best users in terms of the  experienced average throughput. It can be seen that FTGS is indeed able to provide similar throughput to all the users in the system, irrespective of their $\bar \gamma$. PFS, on the other hand, exhibits a larger gap between the average throughput experienced by the different UEs, with a clear penalization of the UEs with worse channel conditions. 

Another aspect of interest is the inter-scheduling time at MAC layer of an UE, which is here defined as the time interval between two consecutive scheduling instants of the UE. In \cite{enhanced}, authors have shown that the inter-scheduling time at the MAC layer can have adverse effects on the TCP congestion control mechanism. Furthermore, inter-scheduling time is related to the delay experienced by users that try to access the channel, and can have a strong impact on real time applications, where delay plays a major role in determining the quality of experience of the final user. 

Since FTGS scheduling decision depends on the channel variations, we argue that in case of slow fading, the inter-scheduling time could be considerably long. The TD approach exhibits higher inter-scheduling time, because all resources in the same TTI are allocated to the same user. In FD approach, instead, different users can be allocated different RBs in the same TTI, leading, on average, to shorter inter-scheduling time, but also smaller transmit capacity at each scheduling event. It is therefore of interest to evaluate the inter-scheduling time in different channel conditions. 

%\begin{figure}[t]
%	\centering		 \includegraphics[width=0.95\columnwidth]{interScheduling.PNG}
%	\caption{Example of a resource allocation pattern for the computation of the inter-scheduling time.}
%	\label{fig:inter_example}
%\end{figure}

The inter-scheduling time has been analyzed assuming the TD approach for FTGS, PFS and BETS, and considering a flat fading Rayleigh channel in both a vehicular ($\DopplerSp=120$ Hz) and a pedestrian ($\DopplerSp=6$ Hz) scenario. We carry out a worst-case analysis by considering the inter-scheduling time of the UE which is scheduled less frequently. Once again, we omit MTS from this comparison since, with this algorithm, the inter-scheduling time of the worst user is much larger than that of the other schedulers. 
 
Let $\delta$ denote the random variable that models the (worst-case) inter-scheduling time for a certain scheduler. From the simulation results, we observed that the empirical statistical distribution of $\delta$ exhibits a peak at the Time Slot duration, i.e., $1$~ms, meaning that the scheduling of each UE occurs in a bursty manner, with runs of slots assigned to the same UEs, followed by periods during which other UEs are served.

\begin{table}[t]
  \centering
    \caption{$P[\delta=1]$ for different schedulers under different fading conditions.}
    \begin{tabular}{|c|c|c|c|c|c|c|}
\cline{2-7}    \multicolumn{1}{c|}{} & \multicolumn{2}{c|}{BETS} & \multicolumn{2}{c|}{FTGS} & \multicolumn{2}{c|}{PFS} \bigstrut\\
    \hline
    $\DopplerSp$ [Hz] & 6     & 120   & 6     & 120   & 6     & 120 \bigstrut\\
    \hline
    $P[\delta=1]$ & 0,044 & 0,032 & 0,956 & 0,501 & 0,210 & 0,356 \bigstrut\\
    \hline
    \end{tabular}%
  \label{tab:delta1}%
\end{table}%

In particular \Tab{tab:delta1} shows $P[\delta=1]$ for the different considered schedulers under a slow and fast fading environment. For FTGS in both environment the probability is significant, from which is clear the attitude of FTGS to allocate resources in a very bursty manner. For BETS the fading environment is not very relevant, and the considered probability is small enough to relate BETS to Round Robin Polling, which always assigns resources to different users in consecutive time intervals. Finally PFS lies in the middle between the two.

\begin{figure}[t]
	\centering		 \includegraphics[width=0.95\columnwidth]{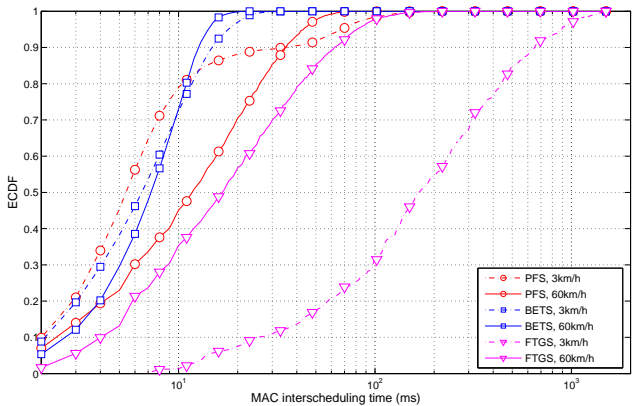}
	\caption{Conditional ECDF of the inter-scheduling time $\delta$, given that $\delta>1$~ms, for the TD version of FTGS ($\triangledown$), BETS ($\Box$) and PFS ($\circ$). Dashed lines:  \emph{slow} flat fading channels ($\DopplerSp=6$~Hz); solid lines: \emph{fast} flat fading channel ($\DopplerSp=120$~Hz).}
	\label{fig:ECDF}
\end{figure}

To appreciate the impact of fading for FTGS, it is hence interesting to investigate the tail of such a distribution. To this end, we report in \Fig{fig:ECDF} the \emph{conditional} empirical cumulative distribution function (ECDF) of $\delta$, given that $\delta>1~$ms. This conditional ECDF captures the statistical distribution of the time between runs of slots allocated to the UE, which is a lower bound of the packet service time at the data link layer (DLL). 

In \Fig{fig:ECDF} we plot results both in the case of slow (dashed-lines) and fast (solid lines) flat fading channels. We see that the conditional statistical distribution of the inter-scheduling time of BETS is not strongly influenced by the dynamics of the fading process, as expected given the channel-agnostic scheduling policy applied by the algorithm. PFS inter-scheduling time exhibits a more pronounced dependence on the fading process, because the scheduling policy also considers the current channel status of UEs. Nonetheless, in most of the cases, $\delta$ does not exceed $110$~ms. The UEs scheduling order imposed by FTGS, instead, is more sensitive to channel variations, so that the inter-scheduling time tail distribution changes quite significantly for fast and slow channels. We can indeed observe that, while with fast fading the FTGS maximum $\delta$ is comparable with that of the other algorithms, with slow fading there is a non negligible probability that $\delta$ exceeds $1$~s. In this case, the packet service time at the MAC layer can sporadically become very large, making this scheduler unsuitable for real time applications. As we will see in the next section, however, the FD version of the scheduler can dramatically improve this performance index in frequency selective channels.

\subsection{Frequency selective channel}
We now turn our attention to frequency selective channels, for which we compare the performance of TD and FD versions of the schedulers to determined whether the increased computational complexity of FD is payed back in terms of significant performance gain or not.

\begin{figure}[t]
	\centering		 \includegraphics[width=0.95\columnwidth]{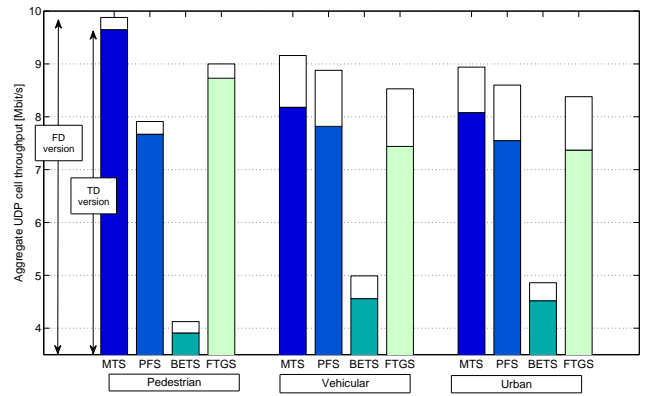}
	\caption{Aggregate cell throughput for the FD (white bar) and TD (coloured bar) versions of MTS, PFS, BETS and FTGS,  with three different frequency selective channels.}
	\label{fig:FD_Rate}
\end{figure}

\begin{figure}[t]
	\centering		 \includegraphics[width=0.95\columnwidth]{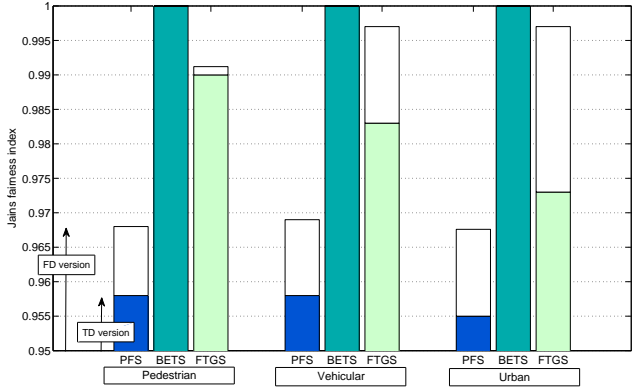}
	\caption{Jain's fairness index for the FD (white bar) and TD (coloured bar) versions of PFS, BETS, and FTGS, with three different frequency selective channels.}
	\label{fig:FD_J}
\end{figure}

\Fig{fig:FD_Rate} shows the aggregate cell throughput achieved by the FD and TD versions of the MTS, PFS, BETS and FTGS, with saturated UDP sources, for the three channel models (Pedestrian, vehicular and Urban) described in detail earlier in \Sec{section:simulation}.  \Fig{fig:FD_J} reports the corresponding fairness index for PFS, BETS and FPGS, while MTS's results are omitted being significantly lower than the others. 

Comparing \Fig{fig:FD_Rate} with  \Fig{fig:rate}, we observe a general throughput loss with respect to the flat fading case, which is only partially compensated by the introduction of the FD version of the algorithms. Furthermore, we note that the performance gap between the FD and TD versions of each scheduler widens for scenarios with higher channel dispersion. 

As expected, the MTS achieves the highest throughput, at the cost of a very low fairness (not reported in the paper). Moreover, we note that %The FD version of PFS shows significant improvement over the TD version, as clearly seen in \Fig{fig:FD_Gain}.  We also note that 
PFS performs better when the channel is more dispersive, consistently with \cite{PFmodelling}. PFS throughput can in fact be expressed as the sum of two terms: the first models the throughput achieved using a Round Robin Scheduler (RRS), while the second is the improvement brought in by the opportunistic approach used by PFS, which is positively correlated to the channel dispersion. PFS, therefore, performs better in severe fading environments. On the other hand, the opportunistic based scheduling policies (such as MTS and FTGS) achieve higher throughput in almost-flat channel environments, where $\tau_{\rm rms}$ is smaller.

%For FTGS, the throughput improvement in the vehicular and urban scenarios is significant ($15$\%) compared to the pedestrian environment (only $3$\%). It is interesting to note that the gain introduced using the FD version of FTGS is always bigger than the one obtained using the FD version of MTS. 

From \Fig{fig:FD_J}, we can see that %shows the fairness results for the PFS, BETS and FTGS schedulers. B
both the TD and FD versions of BETS achieve perfect fairness. Conversely, the fairness of the TD versions of PFS and FTGS degrade for highly dispersive channel, though FTGS still performs better than PFS. 
%For other schedulers, we see that the FD implementation corresponds to a more fair system compared to TD approach, and that FTGS achieves an higher fairness then PFS. Using the TD approach over a very dispersive channel leads to a very poor fairness, both for PFS and FTGS. 
This  loss is likely due to the errors in the channel rate estimate in which the TD versions of the schedulers incur by using the wideband CQI in strongly frequency-selective channels. Conversely, the FD versions of the algorithms make use of the subband CQIs, thus correctly estimating the rate of each single RBG and, hence, better distributing the resources according to the respective utility functions.

The finer granularity in the resource allocation offered by the FD approach also affects the statistical distribution of the inter-scheduling time $\delta$. As noticed in the previous section, this performance index is particularly critical for FTGS, which will hence be the only algorithm considered in the following. While it is easily predictable that the possibility of scheduling multiple UEs in the same time slot will generally reduce $\delta$, and compact its ECDF, the effect of FD on the DLL packet service time is less obvious, because the increased scheduling frequency of each UE comes together with a more fractioned amount of allocated resources. 

To shed some light on these aspects, we introduce the DLL service time $D_i$, which is defined as the time that the $i$-th UE takes to complete the transmission of the head-of-line DLL packet.

\begin{figure}[t]
	\centering		 \includegraphics[width=\columnwidth]{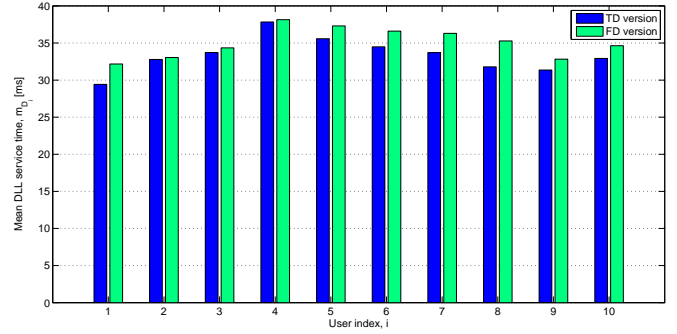}
	\caption{$m_{_{D_i}}$ for the TD and FD approach, over a time dispersive channel with Doppler spread 120 Hz.}
	\label{fig:average_delay}
\end{figure}

\begin{figure}[t]
	\centering		 \includegraphics[width=\columnwidth]{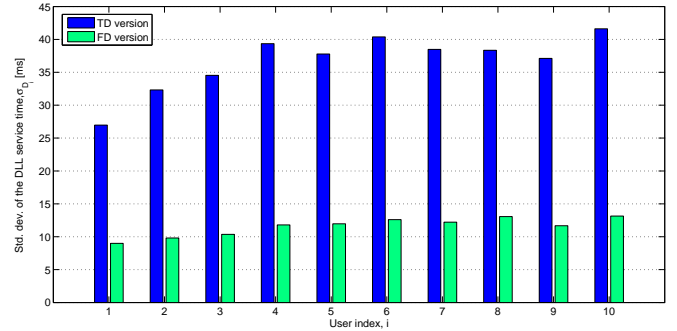}
	\caption{$\sigma_{_{D_i}}$ for the TD and FD approach, over a time dispersive channel with Doppler spread 120 Hz.}
	\label{fig:variance_delay}
\end{figure}

 In Appendix~\ref{DLL} we derive approximate expressions of the mean $m_{D_i}$ and standard deviation $\sigma_{D_i}$ of $D_i$ as functions of the empirical distribution of the inter-scheduling time $\delta$ and of the amount of bits sent at each scheduling event. \Fig{fig:average_delay} and \Fig{fig:variance_delay} report $m_{D_i}$ and $\sigma_{D_i}$, respectively, for each user $i$, in a vehicular scenario, assuming DLL packets of $L=4096$~bytes. Since FTGS guarantees the same long term throughput to all users, both the TD and FD versions of the scheduler yield approximately the same $m_{D_i}$ for each $i$, irrespective of $\bar{\gamma}_i$. However, we note that the FD version of FTGS dramatically decreases $\sigma_{D_i}$, thus offering a more predictable service time to the upper layers. As a side note, we observe that $\sigma_{D_i}$ is slightly higher for users with better average channel conditions (i.e., larger $\bar \gamma_i$), which are indeed scheduled more rarely.

\subsection{Analysis of the opportunistic  gain of FTGS}

The performance analysis carried out so far shows that FTGS is capable of providing high fairness among the UEs, while opportunistically exploiting channel variations to increase the cell throughput. In this section we attempt to quantify such an opportunistic gain when the users average SINRs are more disparate. As benchmark, we consider the performance of BETS, which guarantees equal long-term throughput to all users, but without considering the \emph{current} rates of the different RBGs in the scheduling policy. The analytical expressions of BETS cell throughput and spectral efficiency are derived in Appendix~\ref{BETS}.

We hence define the \emph{opportunistic gain} as 
\begin{equation}
\phi=\frac{\eta_{\!_{FTGS}}}{\eta_{\!_{BETS}}}-1.
\end{equation}
where $\eta_{\!_{FTGS}}$ is the cell spectral efficiency achieved by FTGS, which is obtained from simulations, while  $\eta_{\!_{BETS}}$ is the cell spectral efficiency of BETS as given by \eq{eq:eta_teo_model}.

In all previous results, UEs were located in the LTE cell in order to experience average SINRs in the interval $[10,17.26]~$dB, as reported in \Tab{tab:solution}, with mean cell SINR $\mu_{dB}=15$ dB. We now investigate the FTGS performance when varying the span $\Delta$ of SINRs interval. More precisely, we fix the maximum SINR to $\bar{\gamma}_{max}=25$~dB and progressively decrease the minimum SINR, thus enlarging $\Delta$ and varying the mean cell SINR, $\mu_{dB}$. Note that the SINRs are equally spaced in \emph{linear scale}, thus resulting in a logarithmic distribution over the interval in dB scale.  

%
%We define the SINR spread as
%\begin{equation}
%\Delta=\bar{\gamma}_{max,dB} - \bar{\gamma}_{min,dB},
%\label{eq:Delta}
%\end{equation}
%where $\bar{\gamma}_{max,dB}$ is the average SINR of the best user (in dB) while $\bar{\gamma}_{min,dB}$ is the average SINR of the worst user (in dB). 
%

\begin{figure}[t]
	\centering		 \includegraphics[width=\columnwidth]{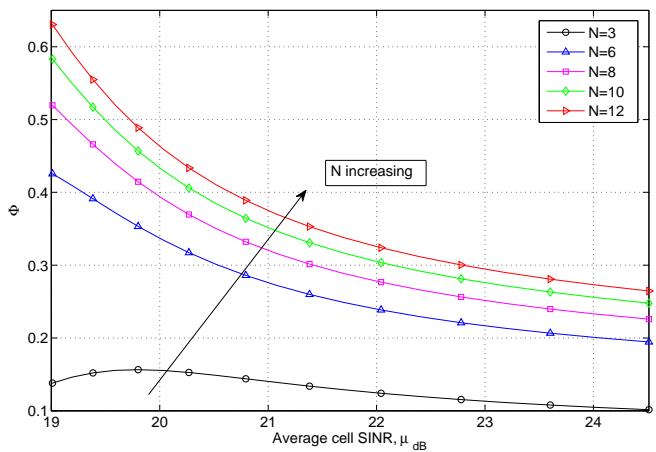}
	\caption{Cell spectral efficiency gain of FTGS over BETS when varying  $\mu_{dB}$, with $\bar{\gamma}_{max}=25$ dB.}
	\label{fig:cellGain}
\end{figure}

In \Fig{fig:cellGain} we report the opportunistic gain $\phi$ when varying $\mu_{dB}$, and for an increasing number of $N$ of UE in the cell. From the figure, it is apparent that the opportunistic gain of FTGS is larger when users average SINRs are more disparate,  since in this condition a channel-aware policy can partially compensate for the worse channel conditions of the more unlucky users. Furthermore, the opportunistic gain increases when the population of users in a given SINR range grows, thus making the opportunistic policies particularly interesting when the number of users is large~\cite{MTS_KandHICC95}.  

%In any case, the presence of a user at the cell edge and, hence, with particularly low average SINR, results in a strong degradation of the overall cell performance. To overcome this problem, the algorithm can be adapted to provide different throughput guarantees according to the average SINRs of the users, though this generalization is left to future work. We can, for example, offer lower throughput guarantees to users with bad channel conditions, and assign the remaining resources to the better users. This would result in a higher cell efficiency while guaranteeing a minimum level of service to users with bad channel conditions.

%\begin{figure}[t]
%	\centering		 \includegraphics[width=\columnwidth]{rangeCell.eps}
%	\caption{Cell spectral efficiency for FTGS (dashed) and the theoretical model (solid), considering a different $\Delta$, with average SINR $\vartheta_{dB}=15$ dB.}
%	\label{fig:rangeCell}
%\end{figure}
%
%
%\begin{figure}[t]
%	\centering		 \includegraphics[width=\columnwidth]{cellMin.eps}
%	\caption{Cell spectral efficiency for FTGS (dashed) and the theoretical model (solid), considering a different $\vartheta_{dB}$, with $\bar{\gamma}_{max}=25$ dB.}
%	\label{fig:cellMin}
%\end{figure}

\subsection{Robustness of FTGS to imperfect channel estimation}
%So far, we have assumed that eNodeB has perfect knowledge of the users' SINR on the different subchannels. This information can be obtained by from control signaling, but to reduce the computational cost and delay, this estimation should be based on a small set of feedback values. This could result in an imperfect estimation. It is of utmost importance, therefore, to evaluate the impact of wrong estimation on the system performance. 
Throughout this work we have assumed that the eNodeB computes the FTGS parameters under the assumption that the different signals are affected by Rayleigh fading. 
To evaluate the robustness of FTGS, we assume that this estimation was wrong and different signals are in fact affected by Rician fading. It is interesting to evaluate the performance loss incurred by FTGS under this scenario\footnote{Actually, the algorithm can be adjusted to other fading distributions but, in this case, the eNodeB shall be able to estimate the most suitable statistical model for the channel from the CQI values returned by the UEs, which is an error-prone and time/resource consuming process.}.  

%We consider a scenario where, based on its SINR estimation, the eNodeB assumes the SINR to be Rayleigh distributed. The optimization problem is also carried out under this assumption. We then simulate the system using a Rician fading channel, and analyze the effect of wrong estimation. 

We hence generated two new fading traces, named Rice1 and Rice2, using the vehicular power delay profile described in Tab.~\ref{tab:pdp}, but adding a strong line-of-sight (LOS) component in the first path for Rice1, and in the first, second and third paths for Rice2. The remaining paths were still assumed to be affected by Rayleigh fading. The Rice factors of the paths affected by Rician fading were set to $K_1=20$~dB in Rice1, and $K_1=10$~dB, $K_2=K_3=0$~dB in Rice2.  

\begin{figure}[t]
	\centering		 \includegraphics[width=0.95\columnwidth]{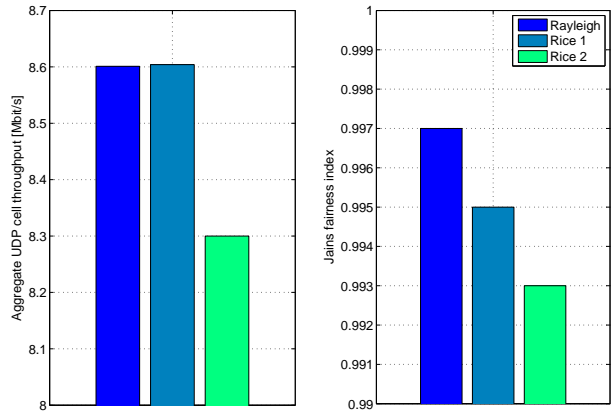}
	\caption{Cell throughput and Jain fairness index in case of Rayleigh and Rice channels.}
	\label{fig:rice}
\end{figure}

\Fig{fig:rice} shows the cell aggregate throughput and the Jain's fairness index for the three channel models, namely Rayleigh, Rice1 and Rice2. We can see that the performance loss of FTGS in presence of strong LOS components in the received signals is insignificant in terms of fairness, and quite limited for the throughput, so that we can conclude that the scheduler is rather robust to different fading models.

\section{Conclusions and Future works}
\label{section:conclusions}

In this work, a QoS-oriented scheduler (the Fair Throughput Guarantee Scheduler (FTGS)) which provides throughput guarantees to different users has been implemented and tested on LTE networks.
A performance analysis is then carried out for the full buffer UDP and TCP traffic sources under both flat and frequency selective fading.
It is shown, by comparing it with the existing well known schedulers, that if the scheduler has knowledge of the users' SINR distribution, it can perform the scheduling decision increasing the system fairness without loosing too much in terms of throughput. The perfect knowledge of the SINR distribution is a strong assumption (although it could be estimated using the CQI feedback), and a dynamic implementation using a real time channel estimation could be subject for future studies. Anyhow we have shown that the scheduler is robust to a certain degree against the wrong estimations of the SINR distribution. 
In addition to implementing the time domain version of the FTGS in \ns, a frequency domain extension of the algorithm has also been proposed and implemented, showing a significant increase of the performance both in terms of throughput and fairness for dispersive frequency selective channels. 
The impact of the scheduling algorithm so chosen on the delay sensitive applications is also studied by analyzing the inter-scheduling time. It is observed that the FD version of the FTGS can substantially reduce the inter-scheduling as compared to the TD version.
Finally the impact of the SINR distribution on the overall system performance has also  been analyzed, showing that the algorithm can be sub-optimal if equal throughput guarantees are promised also to the users experiencing low average SINR. The equal throughput guarantee scheduler hence needs to be tuned for such a scenario and is the subject matter for future work.

\section*{APPENDIX}  % use *-form to suppress numbering

\renewcommand{\thesection}{A-\Roman{section}}
\setcounter{section}{0}  % reset counter 
\renewcommand{\theequation}{A-\arabic{equation}}
  % redefine the command that creates the equation no.
\setcounter{equation}{0}  % reset counter 
 
 \section{Optimization problem derivation}
 \label{Optimization}
 
The aim of the TD version of FTGS is to provide the same long-term throughput to all UEs, while exploiting the channel variability to increase the aggregate cell throughput. Then, considering an arbitrarily long time interval $T_W$, the amount of bits that each UE shall be able to receive can be expressed as
\begin{equation}
	B=T_i \bar{R}_i,
	\label{1}
\end{equation}
where $T_i$ is the time allocated to user $i$ in the time window $T_W$, and $\bar{R_i}$ is the average rate experienced by user $i$ when it is scheduled. Furthermore, assuming that the system is \emph{work conserving}, we require 
\begin{equation}
	\sum_{i=1}^{N}T_i=T_W\,.
	\label{2}
\end{equation}
Denoting by $p(i)$ the access probability of the $i$-th UE, i.e., the fraction of the time user $i$ is scheduled within the time window, \eq{1} can also be expressed as 
\begin{equation}
B=p(i)T_W\bar{R}_i\,.
\label{3}
\end{equation}
The objective is then to find the scheduling probabilities $\{p(i)\}$ for which $B$ is maximized. We observe that, by combining \eq{1}-\eq{3}, we get
\begin{equation}
p(i)=\frac{1}{\bar{R_i}\sum_{j=1}^{N} \bar{R_j}^{-1}}\,,
\label{first}
\end{equation}
so that we need first to find a proper expression for $\bar{R}_i$.

The average rate experienced by user $i$ when scheduled can then be expressed as follows, 
\begin{equation}
\bar{R}_i=W \int_{0}^{\infty}\log_2\Big(1+\frac{\gamma}{\Gamma}\Big)p_\gamma(\gamma|i)d\gamma,
\label{shannon}
\end{equation}
where $p_\gamma(\gamma|i)$ is the probability density function (PDF) of the SINR given that user $i$ is scheduled, and $W$ is the available bandwidth.
%We can now formulate the optimization problem that aims at maximizing the users' throughput in the time window $T_W$, in terms of the scheduling probabilities $\{p(i)\}$. We have: 
%\begin{equation*}
%\max_{p(i)} B,
%\end{equation*}
%such that for all $i=1,2,\dots,N,$
%\begin{equation}
%\bar{R}_i=W \int_{0}^{\infty}\log_2\Big(1+\frac{\gamma}{\Gamma}\Big)p_\gamma(\gamma|i)d\gamma,
%\label{shannon}
%\end{equation}
%where $p_\gamma(\gamma|i)$ is the probability density function (PDF) of the SINR given that user $i$ is scheduled, and $W$ is the available bandwidth.
Assuming a Rayleigh fading channel, and denoting by $\bar{\gamma}_i$ the average SINR for user $i$, the cumulative distribution function (CDF) of the channel SINR is given by
\begin{equation}
P_{\gamma_i}(x) = P[\gamma_i \leq x]=1-\-e^{-x / \bar{\gamma_i}}.
\label{exp}
\end{equation}

We now define a new random variable (r.v.) $S_i \triangleq R_i/\alpha_i$ where $R_i$ is the r.v. that describes the instantaneous rate of user $i$. Therefore, $S_i$ models the priority metric used by the algorithm, as for \eq{metric}. 
The maximum value of $S_i$ with SINR $\gamma$ is given by
\begin{equation}
S_i(\gamma)=\frac{W \log_2\Big(  1+\frac{\gamma}{\Gamma} \Big)}{\alpha_i}.
\label{metric2}
\end{equation}
The CDF of $S_i$ using \eq{exp} is given by
\begin{equation}
	P_{S_i}(s)=P[S_i(\gamma)<s]=1-e^{c / \bar{\gamma}},
	\label{CDFS}
\end{equation}
where $c=\Gamma\big(1-2^{\frac{\alpha_i s}{W}}\big)$. The corresponding PDF is then equal to 
\begin{equation}
p_{S_i(s)}=\frac{\ln(2)\alpha_i}{W \bar{\gamma_i}} (\Gamma-c) e^{c / \bar{\gamma}}. 
\end{equation}

Since $S_i$ is the scheduling metric of FTGS, we can express the access probability of user $i$ as
\begin{equation}
p(i)=P\Big[S_i > \operatorname*{max}_{j \neq i} S_j\Big]=\int_{0}^{\infty}p_{S_i}(s)\prod_{j=1 ,j \neq i}^{N} P_{S_j}(s) {\rm d}s.
\label{second}
\end{equation}
Now, using Bayes' rule, we get the PDF of $S_i$ given that user $i$ is scheduled as
\begin{equation}
p_{S_i}(s|i)=\frac{p_{S_i}(s)}{p(i)}\prod_{j=1 ,j \neq i}^{N} P_{S_j}(s).
\end{equation}
The conditional expectation of $S_i$ given $i$ is then equal to
\begin{equation}
\bar{R_i}=\alpha_i\int_{0}^{\infty}sp_{S_i}(s|i)ds,
\end{equation}
from which we obtain
\begin{equation}
p(i)=\frac{\alpha_i}{\bar{R}_i}\int_{0}^{\infty}s \cdot p_{S_i}(s)\prod_{j=1 ,j \neq i}^{N} P_{S_j}(s){\rm d}s.
\label{third}
\end{equation}

By combining \eq{first}, \eq{second}, and \eq{third} we finally get a set of $3N$ independent equations in $3N$ unknowns, namely $\{\alpha_i,\bar{R}_i,p(i)\}$,  $i=1,2,...,N$, which can be solved using standard numerical tools \cite{opportunistic}.

\section{DLL packet service time statistics}\label{DLL}
We here derive approximate expressions for the first and second order moments of the data link layer (DLL) service time $D$ for a packet of $L$ bits, as functions of the empirical mean and variance of the inter-scheduling time $\delta_k$, and of the number of bits $b_k$ transmitted by the tagged UE at the $k$th scheduling event. The statistical or empirical mean and variance of the generic random variable $x$ will be denoted by $m_x$ and $\sigma_x^2$, respectively. 

Let $N$ denote the number of scheduling events taken by the UE to transmit the DLL packet. We have
$
Y=\sum_{k=1}^{N} b_{k} \geq L\,.
$
For simplicity, we assume $\{b_{k}\}$ are independent and identically distributed random variables. We hence have 
\begin{equation}
m_Y = m_N m_b\,,\quad\hbox{and}\quad \sigma_Y^2=m_N \sigma_b^2 +\sigma_N^2 m_b^2
\label{mysy1}
\end{equation}
Let $P$ denote the bits allocated to the UE in excess of $L$, so that $Y=L+P$. We model $P$ as a random variable uniformly distributed in $[0,b)$. Thus, we get 
\begin{equation}
m_Y = L + m_P = L+\frac{m_b}{2}\,, \quad \hbox{and}\quad \sigma_Y^2 = \sigma_P^2 = \frac{m_b^2+4\sigma_b^2}{12}\,.
\label{mysy2}
\end{equation}
Replacing \eq{mysy2} into \eq{mysy1} we get
\begin{equation}
m_N = L+0.5\quad\hbox{and}\quad \sigma_N^2 =  \frac{m_b^2+4\sigma_b^2}{12\,m_b^2 } - \frac{m_N \sigma_b^2}{m_b^2 }\,.
\end{equation}
Now, the packet service time $D$ can then be expressed as
$
D = \sum_{k=1}^{N} \delta_k
$
from which 
\begin{equation}
m_{D} = m_N m_{\delta}\,,\quad\hbox{and}\quad
\sigma^2_D = m_N \sigma^2_\delta+\sigma_N^2 m_\delta^2\,.
\label{last}
\end{equation}
Using \eq{mysy2} into \eq{last} we get an estimate of the mean and variance of $D$ in terms of the empirical mean and variance of $\delta$ and $b$.

\section{Long-term throughput of BETS}\label{BETS}

We consider $N$ users with average SINR $\lbrace\bar{\gamma}_i\rbrace$, $i=1,2,...,N$. Furthermore, we assume that the received signals are affected by independent Rayleigh fading processes, so that the SINR experienced by user $i$ on any RBG can be expressed as $\gamma_i =\eps \bar \gamma_i $, where $\eps$ is an exponentially distributed random variable of unit mean. Since the resource allocation criterion of BETS does not account for $\gamma_i$, the average rate experience by user $i$ any time it gets scheduled can be expressed as 
\begin{equation}
\bar{G}_i= \mathcal{B} \E{\log_2\Bigg(1+\frac{\eps \bar{\gamma}_i  }{\Gamma} \Bigg)} = \mathcal{B}\log_2(e) e^{\frac{\Gamma}{\bar \gamma_i}}\mathop{E_i}\left(\frac{\Gamma}{\bar \gamma_i}\right) 
\label{eq:G_i}
\end{equation}
where $\mathcal{B}$ is the bandwidth of each RBG and $\mathop{E_i}(x) =\int_{x}^{\infty}(e^{-t}/t)\mathrm{d}t$ is the exponential integral function. 

For ease of explanation, we consider the TD version of the scheduler, though the reasoning can be straightforwardly extended to the FD versions. Since BETS is designed to provide long-term fairness, in a sufficiently long time interval $T$, all users will transmit an equal amount of bits $B$. Hence, the total time allotted to user $i$ in the time window $T$ will hence be equal to $T_i = B/\bar G_i$ for all $i\in\{1,\ldots,N\}$. Therefore, we get
\begin{equation}
T =\sum_{i=1}^{N} T_i=\sum_{i=1}^{N} \frac{B}{\bar{G}_i}\,.
\end{equation}
from which we obtain
\begin{equation}
B=\frac{T}{\sum_{i=1}^{N} ( 1/\bar{G}_i )}.
\end{equation}
Finally, the cell throughput can be computed as 
\begin{equation}
S_{\!_{BETS}}=\dfrac{N \, B}{T} = \frac{N}{\sum_{i=1}^{N} ( 1/\bar{G}_i )},
\end{equation}
and the corresponding cell spectral efficiency is
\begin{equation}
\eta_{\!_{BETS}}=\dfrac{S}{\mathcal{B}}=\dfrac{N\log_2(e)}{\sum_{i=1}^{N}e^{-\frac{\Gamma}{\bar \gamma_i}}\big[ \mathop{E_i}\left(-\frac{\Gamma}{\bar \gamma_i}\right) \big]^{-1}}.
\label{eq:eta_teo_model}
\end{equation}

%
%
%The average number of slots required to transmit the packet can be approximated as
%\begin{equation}
%E[N_i]\simeq L / \bar{b}_i,
%\end{equation}
%and the average packet delay can be computed as
%\begin{equation}
%E[D_i]=E[N_i]E[\delta]
%\end{equation}
%where $E[\delta]$ is the average inter-scheduling time for user $i$.
%The variance of the random variable $D_i$ can be obtained assuming the independence between $N_i$ and $\delta$ and using the random sum variance formula, that yields
%\begin{equation}
%Var[D_i]=E[N_i]Var[\delta]+Var[N_i]E[\delta]^2.
%\label{eq:variance_delay}
%\end{equation}
%In \eq{eq:variance_delay} the only term missing is $Var[N_i]$. Since $E[W_i]=\bar{b}_i/2$ and $E[W_i^2]=E[b_i^2]/3$ we have
%\begin{equation}
%E[Y_i^2]=L^2+L \bar{b}_i + \frac{E[b_i^2]}{3},
%\end{equation}
%so we can finally express $Var[N_i]$ as
%\begin{equation}
%\frac{L^2+L \bar{b}_i + (Var(b_i)+\bar{b}_i^2)/3-E[N_i]Var(b_i)-\Big( E[N_i]\bar{b}_i \Big)^2}{\bar{b}_i^2}
%\end{equation}

\ifCLASSOPTIONcaptionsoff
  \newpage
\fi

\bibliographystyle{IEEEtran}
\bibliography{bibliografia}

\end{document}